\documentclass[osajnl,twocolumn,showpacs,superscriptaddress,10pt]{revtex4-1} 
\usepackage{amsmath,amssymb,graphicx}
\usepackage{upgreek}

\begin{document}

\newcommand{\degree}{^\circ}
\newcommand{\unit}[1]{\,\mathrm{#1}}
\newcommand{\Figureref}[2][]{Figure~\ref{#2}#1}
\newcommand{\Eqnref}[1]{Equation~(\ref{#1})}
\newcommand{\bra}[1]{\left<\mathrm{#1}\right|}
\newcommand{\ket}[1]{\left|\mathrm{#1}\right>}

\newcommand{\affilOU}{Homer L. Dodge Department of Physics and Astronomy, The University of Oklahoma, 440 W. Brooks St. Norman, OK 73019, USA}
\newcommand{\affilStutt}{5. Physikalisches Institut, Universit\"{a}t Stuttgart, Pfaffenwaldring 57 D-70550 Stuttgart, Germany}

\title{Sub-wavelength microwave electric field imaging using Rydberg atoms inside atomic vapor cells}
\date{}
\author{H.Q. Fan}
\author{S. Kumar}
\affiliation{\affilOU}
\author{R. Daschner}
\affiliation{\affilStutt}

\author{H. K\"ubler}
\affiliation{\affilOU}
\affiliation{\affilStutt}
\author{J. P. Shaffer}\email{Corresponding author: shaffer@nhn.ou.edu}
\affiliation{\affilOU}

\begin{abstract}
We have recently shown that Alkali atoms contained in a vapor cell can serve as a highly accurate standard for microwave electric field strength as well as polarization using the principles of Rydberg atom electromagnetically induced transparency. Here, we show, for the first time, that Rydberg atom electromagnetically induced transparency can be used to image microwave electric fields with unprecedented precision. The spatial resolution of the method is far into the sub-wavelength regime. The electric field resolutions are similar to those we have demonstrated in our prior experiments. Our experimental results agree with finite element calculations of test electric field patterns.
\end{abstract}

\maketitle

Atomic standards are important because they enable stable and uniform measurements and often link physical quantities to each other via universal constants \cite{Hall2006}. We have demonstrated in our prior work that atoms contained in a vapor cell can be used for a practical and, in principle, portable microwave (MW) electric field standard using Rydberg atom electromagnetically induced transparency (EIT) \cite{MWpaper,MWVectorPaper}.  The accurate measurement of MW electric field strength and polarization can lead to advances in applications such as antenna design, device development, characterization of electro-magnetic interference, advanced radar applications and materials characterization \cite{Kanda93,Kanda94,Tishchenko03a,Tishchenko03b,camparo1998,swan2001precision}, including metamaterials \cite{MetaMaterialsNearField1,MetaMaterialsNearField2,MetaMaterialsOpt}.

To our knowledge, no other work exists on imaging MW electric fields with atoms in vapor cells. Even in the field of magnetometry, where vapor cell magnetometers have played a central part \cite{Budker2007}, absorption imaging for vapor cell MW magnetometry has only been recently reported \cite{TreutleinMWhot,horsley2013imaging}. Many of the technical issues of imaging a MW magnetic field as opposed to an electric field with a vapor cell are different. Knowledge of both fields is important. Despite the rather straightforward connection between the electric and magnetic fields in free space, there is not always a simple relation between them in the near field. The absolute measurement of MW electric fields at sub-wavelength resolutions and in the near field is necessary for many MW applications.

To meet the need for sub-wavelength imaging of MW electric fields, we demonstrate a scheme for sub-wavelength MW electrometry using Rydberg atom EIT \cite{Fleisch05,acell} in Cesium (Cs) atomic vapor cells at room temperature. In contrast to scanning probe technology \cite{ScanningProbeMW,dutta1999imaging}, our approach avoids cryogenics and eliminates the presence of conducting materials near the sample, therefore minimizing field disturbances. We achieve a 2-dimensional spatial resolution of $\sim \lambda_{\mathrm{MW}}/650$, $\sim 66\,\mu$m at $\sim 6.9\,$GHz, using a test MW electric field in the form of a standing wave and image the MW electric field directly above a co-planar waveguide (CPW) to demonstrate near field imaging. The electric field resolution is $\sim 50\, \mu$V$\,$cm$^{-1}$ limited by our detection setup. The measurements are compatible with our prior work where we attained a minimum detectable electric field amplitude of $\sim 8\,\mu \mathrm{V}\,$cm$^{-1}$ and a sensitivity of $\sim 30 \,\mu\mathrm{V}\,$cm$^{-1}\,$Hz$^{-1/2}$ \cite{MWpaper} with an angular resolution of $0.5\degree$ in vector measurements \cite{MWVectorPaper}.

\begin{figure}[htbp]
\includegraphics[width=\columnwidth]{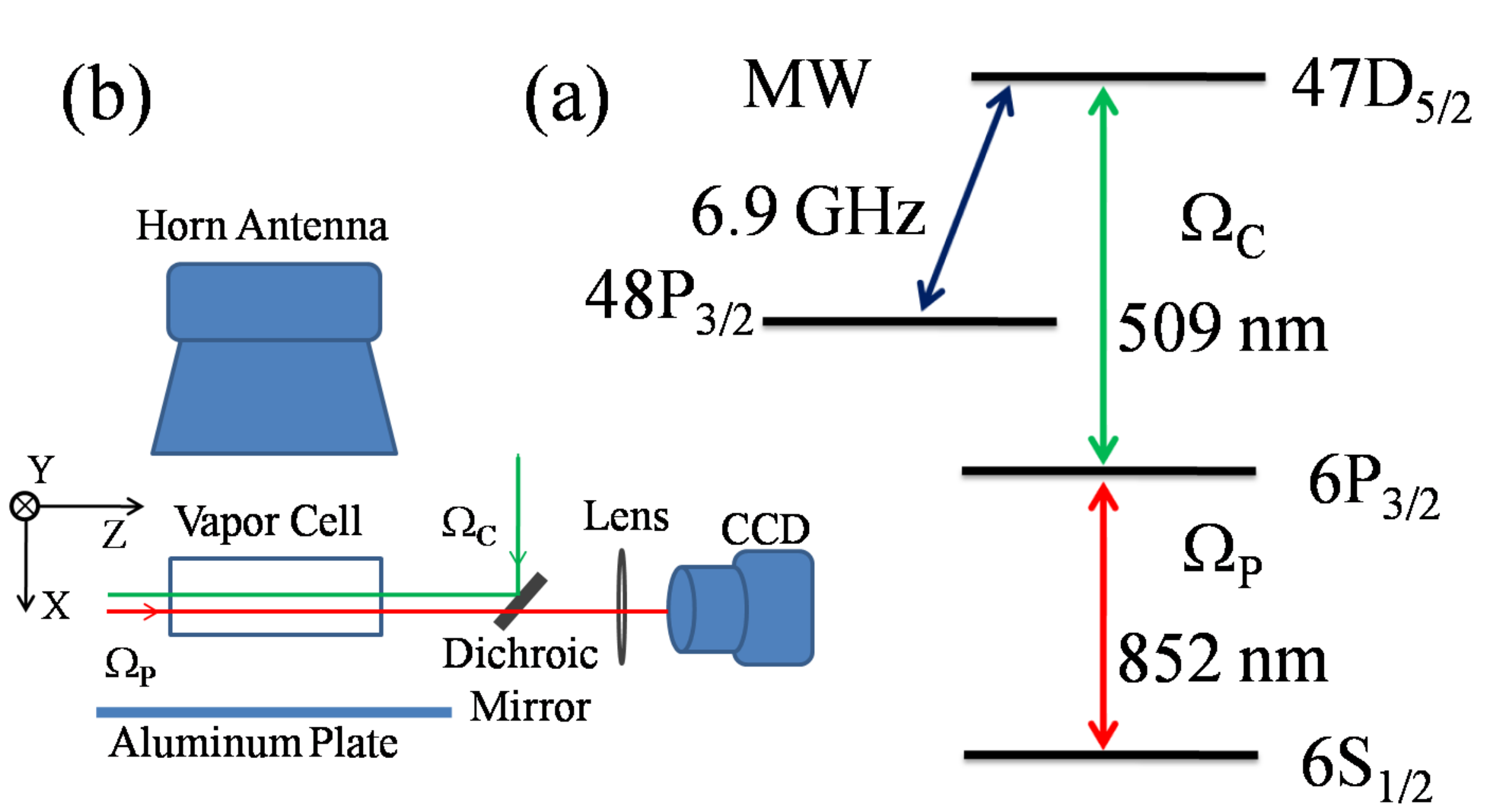}
\caption[short caption]{(color online) Setup and energy level diagram for the experiments. (a) $\Omega_\mathrm{p}$ and $\Omega_\mathrm{c}$ are the EIT probe  and coupling transition Rabi frequencies, respectively. MW labels the transition used for the standing wave electric field measurements. (b) The MWs supplied by the horn antenna form a standing wave between the aluminum plate and antenna that is sampled by the atoms in the Cs vapor cell. The horn antenna and plate are moved together to change the position of the vapor cell in the standing wave MW electric field.\label{fig:1}}
\end{figure}

To measure the MW electric field strength of a standing wave field at $\sim 6.9\,$GHz, we use the Cs level system shown in Fig.~\ref{fig:1}a. In the 3-level EIT system, $\mathrm{6S_{1/2}-6P_{3/2}-47D_{5/2}}$, a ``dark state'' is created by a probe, $\Omega_\mathrm{P}$, and a coupling laser, $\Omega_\mathrm{C}$, such that resonant absorption of the probe laser is prohibited. Coupling a fourth level to this EIT system, $\mathrm{48P_{3/2}}$, with a MW electric field can create a ``bright state'' that causes probe photons to again be absorbed on resonance \cite{MWpaper,Dutta07,imamoglu1996,Lukin1999}. The bright state induced by the MW electric field can appear in the probe absorption spectrum as an Autler-Townes splitting of the 3-level EIT dark state for large enough MW electric amplitudes \cite{MWpaper}. The splitting of the spectrum is proportional to the Rabi frequency, $\Omega_{\mathrm{MW}} = \mu_{\mathrm{MW}}\, E_{\mathrm{MW}}/ \hbar$, where $E_{\mathrm{MW}}$ is the MW electric field amplitude and $\mu_{\mathrm{MW}}$ is the transition dipole moment for the $\mathrm{47D_{5/2}}\longleftrightarrow\mathrm{48P_{3/2}}$ transition. The sensitivity to the MW electric field is large because $\mu_{\mathrm{MW}}$ is large, $> 1000\,$e$\,$a$_0$. By scanning the probe laser frequency and imaging the spatially dependent absorption with a CCD camera, the amplitude of the MW electric field can be acquired across the overlap region of the probe and coupling laser beams \cite{MWpaper}. In the experiments we present here, the probe and coupling lasers have opposite circular polarizations and we measure the amplitude of the MW electric field \cite{MWVectorPaper}. The setup is shown in Fig.~\ref{fig:1}b.

The probe laser light that drives the $\mathrm{6S_{1/2} (F = 4) \rightarrow 6P_{3/2} (F= 5)}$ transition at $852\,$nm is generated by a frequency stabilized diode laser \cite{Zeemanpaper}. The coupling laser at $509\,$nm is supplied by an amplified diode laser at $1018\,$nm that is doubled in a ring cavity. The coupling laser is frequency stabilized and tuned to the $\mathrm{6P_{3/2} \rightarrow 47D_{5/2}}$ transition of Cs. The probe laser power is 50$\,\mu$W resulting in $\Omega_\mathrm{P} = 2.033\,$MHz. The coupling laser power is $34\,$mW corresponding to $\Omega_\mathrm{C} = 1.124\,$MHz. The diameter of the probe (coupling) laser beam is $2.78\pm 0.01\,$mm ($2.47\pm 0.01\,$mm). $\mu_{\mathrm{MW}}= 2938.5 \,$e$\,$a$_0$ which is 1000 times larger than the Cs $\mathrm{D2}$ transition dipole moment \cite{SteckCs}. The probe and coupling lasers are setup in a counter-propagating geometry. An HP 8340B is used for the MW source.

$\Omega_\mathrm{P}$ and $\Omega_\mathrm{C}$ are chosen to reduce the Rydberg state population, yet still obtain signals that facilitate the measurement in a reasonably large spatial area. Rydberg state population results in Rydberg atom collisions and absorption of MWs as they pass through the laser beam overlap region. Absorption can be accounted for using Beer's law for the MW powers in our experiments. Our Rabi frequencies, MW electric field amplitudes and gas densities make collisions and MW absorption negligible. These considerations and the constraints imposed by the limited coupling laser power result in the measurements taking place in the intermediate regime of EIT, outside the weak probe limit \cite{Fleisch05}. The qualitative description of the method given is unchanged.

\begin{figure}[htbp]
\includegraphics[width=\columnwidth]{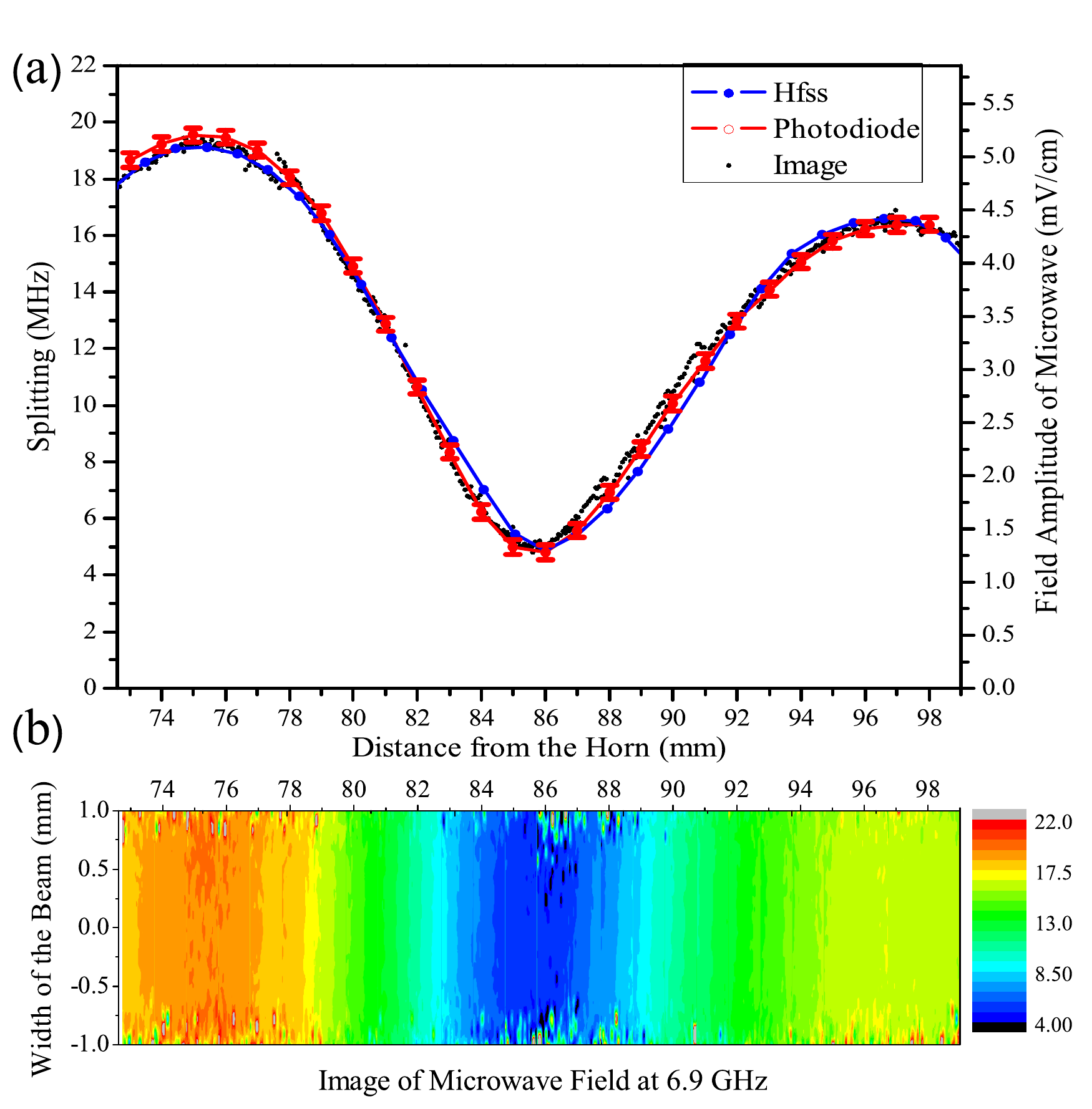}
\caption[short caption]{(color online) (a) Measurement of the EIT dark state splitting versus distance from the horn antenna. The theoretical FE calculation (blue), photodiode measurement (red) and image measurement data (black) are shown for a trace taken along the centerline of the image. The right hand axis shows the MW electric field amplitude that corresponds to the EIT dark state splitting (b) Two-dimensional image of the MW electric field. The data is composed of 26 images with a step size of $1\,$mm in the direction along the cavity axis, x. In the transverse, y, direction, a $2\,$mm cross-section with the zero point at the center of the probe beam is displayed. The legend shows the EIT dark state splitting in MHz. The error bars show the standard deviation between the photodiode and imaging measurements. \label{fig:2}}
\end{figure}

Two room temperature Cs vapor cells were used to show that the measurements do not depend on the vapor cell geometry within the uncertainty of our experiments. One vapor cell was quartz with a square cross-section, $1\,$cm$\,\times 1\,$cm$\,\times 3\,$cm, while a second vapor cell was pyrex with a cylindrical cross-section, $2.5\,$cm diameter and $4\,$cm long. Each cell was evacuated to $< 3 \times 10^{-9}\,$Torr before it was filled with Cs. We use three Helmholtz coils to cancel residual magnetic fields. An anechoic box is placed around the setup to attenuate reflections and stray MW electric fields. For all the measurements, the vapor cell is placed in a region of the test MW electric field where the field is maximally uniform along the length of the vapor cell, Fig.~\ref{fig:1}b. We measure two-dimensional MW electric fields.

The probe laser transmission signal is captured by a CCD camera (Pco Pixelfly). The probe beam is imaged on the CCD camera using a lens, $f=400\,$mm. The cell is placed at a position $2f$ in front of the lens while the camera is located a distance $2f$ behind the lens. The depth of field is $\sim 9\,$mm. The diameter of the Airy disk for the imaging system is $\sim 65\,\mu$m.

Images of the probe transmission with the coupling laser on and then off are subtracted to obtain the dark state EIT spectrum at a particular probe laser frequency.  The exposure time is longer than the lifetime of the Rydberg states, $\sim 50\,\mu$s, so that steady-state is reached. We chose an exposure time of 1 ms. Each image consists of $40$ averaged pictures. The spatial dependence of the MW electric field is obtained by averaging the probe laser transmission over small patches of each image. The dimension of the patches was $66\,\mu$m$\times 66\,\mu$m. The MW electric field at each patch is obtained by fitting the probe laser spectrum obtained from a series of images taken at different probe laser frequencies to get the local splitting of the EIT dark state.

The MW electric field pattern we investigated first is created by placing a reflecting aluminum plate $3\lambda_{\mathrm{MW}}$ away from a horn antenna to create a cavity, Fig.~\ref{fig:1}b, in which a standing wave is formed with a period of $\lambda_\mathrm{MW}/2$ as predicted by a finite element (FE) calculation using High Frequency Structural Simulator (Hfss). The horn antenna generates a MW field with wavelength $\lambda_{\mathrm{MW}}=4.3187099$ cm and frequency $6.9465189\,$GHz. To measure the standing wave MW electric field, the horn antenna and reflective plate are fixed on a translation stage so that they can be moved together to scan the atoms in the vapor cell along the electric field pattern. Fig.~\ref{fig:2}a shows the image data extracted along the centerline of the MW electric field pattern compared to the FE  calculation for a spatial resolution of $66\,\mu$m over approximately one half wavelength.  Images are taken every $1\,$mm along the axis of the cavity, x direction, Fig.~\ref{fig:1}b. Good agreement is observed between the two traces. Differences can be attributed to the finite calculation region, cell size, and modeling of the horn antenna used for the FE calculation. Fig.~\ref{fig:2}b shows the image data. The figure is composed of 26 images each offset by $1\,$mm in the x direction. The transverse axis is centered on the y symmetry plane of the cell and cavity, Fig.~\ref{fig:1}b, as can be seen by inspection of Fig.~\ref{fig:2}b.

\begin{figure}[htbp]
\includegraphics[width=\columnwidth]{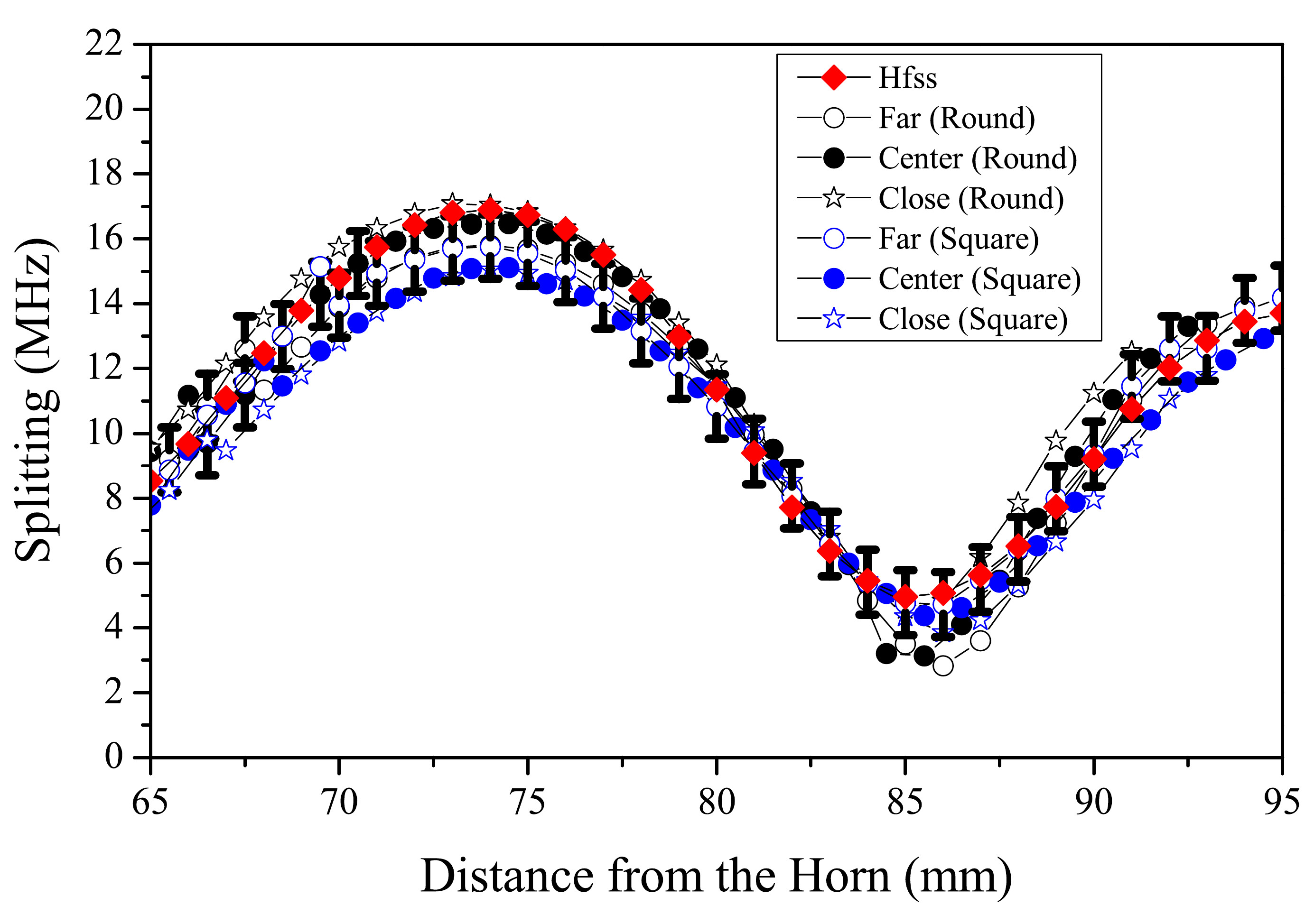}
\caption[short caption]{(color online) Photodiode measurement of the EIT splitting versus distance at different positions of the probe beam for square (blue) and cylindrical (black) cells compared to the FE calculations (red). The representative error bars are the standard deviation of the measurement from the theory. \label{fig:3}}
\end{figure}

As a test, we placed an iris at the center of the probe beam and observed the spectrum of the probe laser with a photodiode, similar to experiments found in Refs.~\cite{MWpaper,MWVectorPaper}. Fig.~\ref{fig:2}a shows this measurement in comparison with the data obtained from the images and FE calculation. By analyzing the difference between the photodiode measurement and the image measurement for different traces along x, the standard deviation is $0.25\,$MHz at a spatial resolution of $66\,\mu$m. This uncertainty translates to $67\,\mu$V$\,$cm$^{-1}$. The associated spatial resolution is $\sim \lambda_\mathrm{MW}/650$. A better spatial resolution can be obtained at the cost of uncertainty in the MW electric field and a reduced depth of field. These values are all limited by our CCD camera and imaging setup, as is the sensitivity. The sensitivity is low because it is limited by the CCD camera readout time. A more sensitive and faster CCD camera, or multichannel spatially sensitive detector, can lead to large improvements. Our proof of principle measurements, while impressive, can be significantly better with more engineering of the detection system.

\begin{figure}[htbp]
\includegraphics[width=\columnwidth]{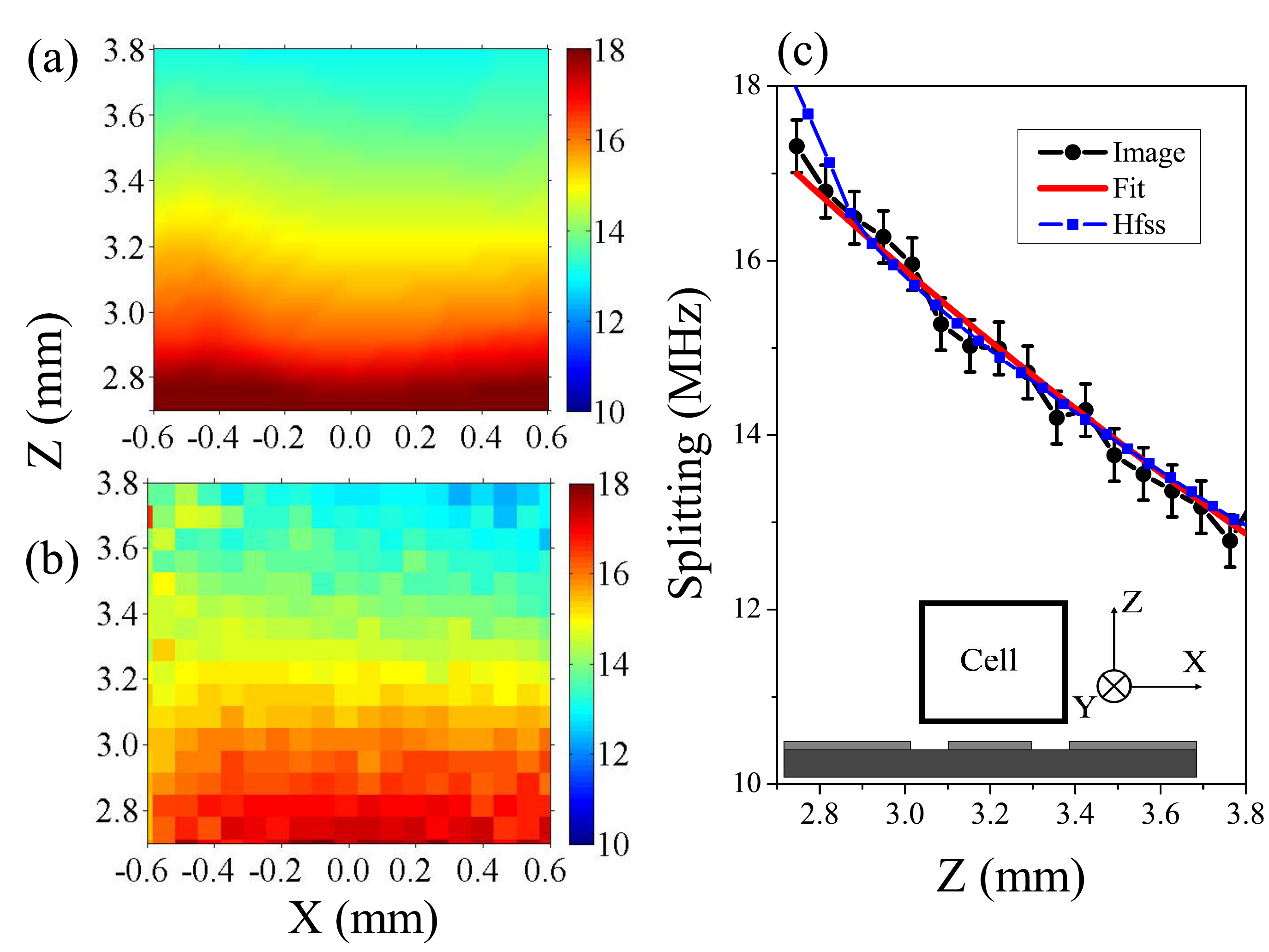}
\caption[short caption]{(color online) (a) This figure shows the FE calculation directly above the microwave CPW where the MW electric field was sampled. (b) This figure shows the imaging data corresponding to (a). (c) This plot shows a comparison between the data and the FE calculation along the $z$-axis at $x = 0$. $x=0$ is at the center of the CPW. Solid circles (black) are the imaging data while the solid squares (blue) are the FE calculation. The line (red) is a plot of $A \,\mathrm{exp}(- 2 \pi z /\lambda_\mathrm{MW})$, where $A$ is a constant determined by a fit to the data. The geometry is shown in the inset of (c).  \label{fig:4}}
\end{figure}

To further check our results, we used an iris in front of a photodiode to filter the probe signal to measure the dark state EIT splitting at three different horizontal positions across the square and cylindrical vapor cells. The probe beam size that was used was $\sim 500\,\mu$m. The spots were spaced horizontally by $1\,$mm along the x-direction, on the centerline of the vapor cells. The data for this measurement is compared to an FE calculation in Fig.~\ref{fig:3}. The results obtained for the two different vapor cells show that the signal, at our uncertainty, is independent of vapor cell size and composition. The fact that the wavelength of the MW electric field is larger than the dimensions of either cell explains this observation to a great extent. The primary observable effect of the vapor cells is an increase in the optical path length of the cavity, which can be observed as a shift in the fringe pattern. This effect has been removed from the graph in Fig.~\ref{fig:3}. Notice that Fig.~\ref{fig:3} shows variation where the low vs. high MW electric field measurements corresponding to the two vapor cells invert themselves over the extent of the scan, suggesting that the MW electric field errors are not correlated with a particular vapor cell.

To show that near field, sub-wavelength imaging is possible, we investigated the MW electric field above a CPW and compared our results to FE calculations of the MW electric field. The CPW was operated at a frequency of $12.602001\,$GHz, $\lambda_\mathrm{MW} = 2.3805743\,$cm. The MW electric field drives the $\mathrm{39D_{5/2} \longleftrightarrow 40P_{3/2}}$ transition. The Rabi frequencies of the probe and coupling lasers were $2.033\,$MHz and $1.516\,$MHz, respectively. The transition dipole moment for the $\mathrm{39D_{5/2} \longleftrightarrow 40P_{3/2}}$ transition is $\mu_{MW} = 1977.1\,$e$\,$a$_0$. Fig.~\ref{fig:4} shows the results along with the FE calculation of the cell above the waveguide. The FE calculation included the cell. From the comparison of the data to the calculation we find a statistical variation in the field to be $120\,\mu$V$\,$cm$^{-1}$. The spatial resolution is $66\,\mu$m.

In conclusion, we presented a method to image MW electric fields using Rydberg atom EIT in a Cs vapor cell at room temperature. By mapping the MW electric field of a standing wave and a near field region above a CPW, very high spatial resolution has been demonstrated at MW electric field resolutions similar to the best so far achieved in absolute measurements. This MW electric field imaging method promises to have a wide range of applications, particularly in the area of characterizing metamaterials and small MW circuits. The method can be improved by using better imaging detectors and smaller vapor cells \cite{MicroCellPaper,Baluktsian10}. It may be possible to measure much smaller MW electric fields at higher sensitivity, determined by the Rydberg state lifetime. Achievement of this objective could allow MW electric field imaging at field strengths $< 10\,$nV$\,$cm$^{-1}$ with $< 10\,\mu$m spatial resolution.

We thank the \emph{Institut f\"{u}r Elektrische und Optische Nachrichtentechnik} and H. Sigmarsson for advice on waveguides, and \emph{Rogers Corp.}, for samples. This work was supported by the DARPA Quasar program by a grant through ARO (60181-PH-DRP), AFOSR (FA9550-12-1-0282) and NSF (PHY-1104424).

\newpage

%

\end{document}